\documentclass[pre, twocolumn, showkeys, amsmath, amssymb]{revtex4}

\usepackage{graphicx}

\begin{document}

\title{Sleep as the solution to an optimization problem}

\author{Emmanuel Tannenbaum}
\email{emanuelt@bgu.ac.il}
\affiliation{Department of Chemistry, Ben-Gurion University of the Negev,
Be'er-Sheva 84105, Israel}

\begin{abstract}

This paper develops a highly simplified model with which to 
analyze the phenomenon of sleep.  Motivated by Crick's suggestion
that sleep is the brain's way of ``taking out the trash,'' a suggestion
that is supported by emerging evidence, we consider the problem of the
filling and emptying of a tank.  At any given time, the tank may take
in external resource, or fill, if resource is available at that time, 
or it may empty.  The filling phases correspond to information input
from the environment, or input of some material in general, while the emptying
phases correspond to processing of the resource.  Given a 
resource-availablility profile over some time interval $ T $, we
develop a canonical algorithm for determining the fill-empty profile
that produces the maximum quantity of processed resource at the end
of the time interval.  From this algorithm, it readily follows that
for a periodically oscillating resource-availability profile, the
optimal fill-empty strategy is given by a fill period when resource
is available, followed by an empty period when resource is not.  
This cycling behavior is analogous to the wake-sleep cycles in
organismal life, where the generally nocturnal sleep phase is a period
where the information collected from the day's activities may be processed.
The sleep cycle is then a mechanism for the organism to process a maximal
amount of information over a daily cycle.  Our model can exhibit 
phenomena analogous to ``microsleeps,'' and other behavior associated
with breakdown in sleep patterns.

\end{abstract}


\maketitle

\section{Introduction}

Although sleep is exhibited by almost all complex multicellular life, it is 
still a largely poorly understood phenomenon.  The difficulty in understanding 
the necessity of sleep derives from the observation that sleep-deprived organisms 
do not show signs of physical damage, and yet they are characterized by impaired 
cognitive abilities.  In extreme cases, sleep deprivation can even lead to death.

A hypothesis for sleep, advanced by Francis Crick, is that sleep is
a way for the brain to perform various upkeep, or alternatively,
``garbage collecting'' functions necessary for proper brain function
\cite{SLEEP1}.  More specifically, sleep is a time when the brain sorts 
through various stored memories, and discards those deemed unessential, while 
processing those deemed essential for long-term storage.

There is now evidence suggesting that Crick's hypothesis may be correct:
It has been discovered that neurons contain a protein, termed Fos,
which is involved in proper neuronal function \cite{SLEEP1}.  During periods of
neuronal stimulation, Fos naturally builds up, apparently as a by-product
of various neuronal activities.  Proper neuronal function is no longer
possible once Fos levels reach a critical level.  During the sleep state of 
an organism, Fos levels rapidly drop.  Apparently, the Fos protein acts as 
a molecular switch that regulates various genes involved in proper neuronal 
function \cite{SLEEP1}.  During sleep, these genes are suppressed, allowing
the neuron to re-set for the next period of wakefulness.

A number of appropriate analogies for understanding the phenomenon of sleep
include the build-up of lactic acid during periods of intense muscular activity,
and the necessity for corporations, or even individuals, to periodically significantly
reduce their level of external interactions and to update various inventories and
finances.

Based on Crick's hypothesis, we argue that the phenomenon of sleep occurs because,
given alternating cycles of available sensory input (day-night), a maximal amount
of information processing occurs when the brain devotes its resources to collecting
information when it is available, and to then process that information when 
external information is significantly less available.

In this paper, we proceed to develop a model for the filling and emptying
of a tank.  Filling the tank corresponds to the input of external information
into a biological system (a brain or even a single neuron), while emptying the
tank corresponds to the processing of the information.  For a given resource 
availability profile over some time period, we develop a canonical fill-empty
profile that we rigorously prove optimizes the total amount of resource processed. 
We then go on to consider specific resource availability profiles and associated
optimal solutions, and draw analogies between the optimal profiles and observed
behavior.

We point out that while mathematical models related to sleep have been previously
developed \cite{SLEEP2, SLEEP3, SLEEP4, SLEEP5}, such models have primarily focused
on the dynamics associated with the underlying pathways controlling the wake-sleep
cycle.  However, the need for such pathways have not been addressed.  

This paper presents a model in which an alternating pattern of resource input and 
processing emerges as the solution to an optimization problem, which in the case 
of sleep presumably leads to a fitness advantage for the organism.  To our knowledge, 
a mathematical model whereby sleep emerges as the solution to an optimization problem has not
been previously advanced.

\section{Separation of Tasks and Optimization of Output}

Before presenting the actual tank filling model in the following section,
we give an example illustrating how separation of tasks can result in
a higher level of information processing than if various tasks are performed
simultaneously.

So, consider a system that is capable of engaging in two tasks:  (1)  An
``input'' task, whereby external resource is input into the system.  (2)
A ``processing'' taks, whereby the resource is processed for use by the system.

If we are dealing with a biological system, then we assume that the function
of each task is mediated by some protein.  For the ``input'' tasks, we denote
the protein by $ P_1 $, while for the ``processing'' tasks, we denote the
relevant protein by $ P_2 $.  Now, $ P_1 $ and $ P_2 $ are encoded in a genome,
with corresponding genes denoted by $ G_1 $ and $ G_2 $.  If we assume a total
transcription plus translation rate of $ k_T $, then when only one gene is active,
the active gene $ G_i $ produces protein at a rate $ k_T $, while when both genes
are active, each protein is produced at a rate $ (1/2) k_T $.  We assume that
at least one of the genes is active at any given time.

Furthermore, we assume that the proteins have a decay rate given by a first-order
constant $ k_d $.  Therefore, if $ \epsilon_1(t) $ is defined to be $ 0 $ if
gene $ G_1 $ is off, and $ 1 $ if gene $ G_1 $ is on, and $ \epsilon_2(t) $ is
defined analogously for gene $ G_2 $, then, letting $ n_{P_1} $, $ n_{P_2} $ 
denote the number of proteins in the system at a given time, we have,
\begin{eqnarray}
&   &
\frac{d n_{P_1}}{dt} = \frac{1}{2} (\epsilon_1(t) - \epsilon_2(t) + 1) k_T - k_d n_{P_1}
\nonumber \\
&   &
\frac{d n_{P_2}}{dt} = \frac{1}{2} (\epsilon_2(t) - \epsilon_1(t) + 1) k_T - k_d n_{P_2}
\end{eqnarray}

Now, if we let $ \delta(t) $ denote a resource availability profile, defined to be $ 1 $
when external resource is available, and $ 0 $ otherwise, then in the simplest assumption
the rate at which resource enters the system is proportional to $ n_{P_1} $, and the
rate at which resource is processed is proportional to $ n_{P_2} $.  If $ n_1 $ denotes
the amount of unprocessed resource at any given time $ t $, and $ n_2 $ denotes the
amount of processed resource, then we have,
\begin{eqnarray}
&   &
\frac{d n_1}{dt} = r_1 n_{P_1} \delta(t) - r_2 n_{P_2} (1 - \delta_{n_1, 0})
\nonumber \\
&   &
\frac{d n_2}{dt} = r_2 n_{P_2} (1 - \delta_{n_1, 0})
\end{eqnarray}
where $ \delta_{n_1, 0} = 1 $ if $ n_1 = 0 $, and $ 0 $ otherwise.

Note then that if $ \delta(t) $ experiences oscillatory periods of resource
availability, then keeping both genes on at all times may not lead to 
an optimal processing of resource.  The reason for this is that when $ \delta(t) = 1 $,
resource only enters the system at half the maximal rate.  Depending on the 
resource availability profile, it may be optimal for the system to only
take in external resource when it is available, and then process that resource during
periods when resource is significantly less available.

We note that for finite values of $ k_T $ and $ k_d $, switching from one
task to another involves a transient during which the proteins involved in one
task degrade and the proteins for the other task reach their steady-state levels.
However, if $ k_T, k_d \rightarrow \infty $ in such a way that $ k_T/k_d $ is
fixed, then there is no time associated with switching tasks.  In this case,
we may assume, for simplicity, that only one task can be active at any given time,
since a profile where both tasks are on over a time interval can be approximated
to arbitrary accuracy by a profile that rapidly oscillates between one task and the
other.

\section{A Tank Filling Model}

\subsection{Model description}

Consider a tank that can be filled and emptied with some unspecified material.
At any given time $ t $, external resource is available, or it is not.  We denote the
resource availability profile by a function $ \delta(t) $, where $ \delta(t) = 1 $
if resource is available at time $ t $, and $ 0 $ if not.  If $ \delta(t) = 1 $, then
the tank may be filled at a rate $ r_f $.  The tank may also be emptied at a rate
$ r_e $ as long as it is not empty.  Over any finite interval $ [t_1, t_2] $, we
assume that $ \delta(t) $ is discontinuous at a finite number of points.  This implies
that $ \delta(t) $ may be taken to be a piecewise constant function.

At any given time $ t $, we assume that the tank is carrying out one of the fill or
empty tasks, but that the tasks cannot occur simulatenously.  We let $ \epsilon_f(t) $
denote the fill profile function, so that $ \epsilon_f(t) = 1 $ if the tank is in the
fill mode at time $ t $, and $ \epsilon_f(t) = 0 $ otherwise.  We let $ \epsilon_e(t) $
denote the empty profile function, so that $ \epsilon_f(t) + \epsilon_e(t) = 1 $ at 
all times.  As with $ \delta(t) $, we assume $ \epsilon_f(t) $, $ \epsilon_e(t) $ are 
piecewise constant, with a finite number of discontinuities over a finite interval.

We also let $ n_T(t) $ denote the total amount of material in the tank at time $ t $,
and $ n_P(t) $ denote the total amount of material that has been processed through
the tank at time $ t $.  It should be apparent that,
\begin{eqnarray}
&   &
\frac{d n_T}{dt} = r_f \epsilon_f(t) \delta(t) - r_e \epsilon_e(t) (1 - \delta_{n_T, 0})
\nonumber \\
&   &
\frac{d n_P}{dt} = r_e \epsilon_e(t) (1 - \delta_{n_T, 0})
\end{eqnarray}

\subsection{Optimal fill-empty profiles}

Given a resource availability profile $ \delta(t) $ over some time interval
$ [0, T] $, we wish to determine the fill-empty profile $ \epsilon = 
(\epsilon_f, \epsilon_e) $ that maximizes $ n_P(T) $, given the initial 
conditions $ n_T(0) = n_P(0) = 0 $.  As a notational convenience,
we let $ n_{\epsilon, T}(t) $, $ n_{\epsilon, P}(t) $ denote the
$ n_T $ and $ n_P $ values associated with the fill-empty profile
$ \epsilon $.

A natural fill-empty profile, denoted $ \epsilon_0 = (\epsilon_{f, 0}, \epsilon_{e, 0}) $,
is defined by the following prescription:  Fill whenever $ \delta(t) = 1 $,
empty whenever $ \delta(t) = 0 $ as long as $ n_T(t) > 0 $.  Continue
with this fill-empty profile until $ n_T(t) = r_e (T - t) $, at which
point the tank should be emptied until time $ T $.  Let $ t_{\epsilon_0} $ 
denote the critical time at which emptying until time $ T $ begins.
Then for notational convenience, we define $ I_1 = [0, t_{\epsilon_0}] $,
$ I_2 = [t_{\epsilon_0}, T] $.

We now prove that $ \epsilon_0 $ yields a maximal value 
for $ n_P(T) $.  We begin by defining, for an arbitrary fill-empty
profile $ \epsilon $ over some set $ {\bf S} $, the quantities 
$ T_f({\bf S}; \epsilon) $, $ T_e({\bf S}; \epsilon) $, $ T_w({\bf S}; \epsilon) $,
as follows:  We define $ {\bf S}_f = \{t \in {\bf S}| \delta(t) = 1 \mbox{ and }
\epsilon_f(t) = 1 $, $ {\bf S}_e = \{t \in {\bf S}| n_{\epsilon, T}(t) > 0 \mbox{ and }
\epsilon_e(t) = 1\} $, and $ {\bf S}_w = {\bf S}/({\bf S}_f \bigcup {\bf S}_e) $.
If $ \mu(\Omega) $ denotes the measure of a set $ \Omega $ (essentially
the total length of the set), then $ T_f({\bf S}; \epsilon) \equiv \mu({\bf S}_f) $, 
$ T_e({\bf S}; \epsilon) \equiv \mu({\bf S}_e) $, and $ T_w({\bf S}; \epsilon) \equiv \mu({\bf S}_w) $.
We should point out that because $ \delta $, $ \epsilon_f $, $ \epsilon_e $ are assumed
to be piecewise constant, all sets considered in this paper are unions of disjoint
intervals, and hence are measurable.

Given a set $ {\bf S} $, define $ {\bf S}^{0} = \{t \in {\bf S}|
\delta(t) = 0\} $, and $ {\bf S}^{1} = \{t \in {\bf S}: \delta(t) = 1\} $.
Then define $ T_{e, 0}({\bf S}; \epsilon) = \mu({\bf S}^{0}_e) $,
and $ T_{e, 1}({\bf S}; \epsilon) = \mu({\bf S}^{1}_e) $.
Note that $ T_e({\bf S}; \epsilon) = T_{e, 0}({\bf S}; \epsilon) +
T_{e, 1}({\bf S}; \epsilon) $.

Note that since $ \epsilon_{0, f}(t) = 1 $ whenever $ \delta(t) = 1 $
for $ t \in I_1 $, it follows that $ T_f(I_1; \epsilon) \leq T_f(I_1; \epsilon_0) - 
T_{e, 1}(I_1; \epsilon) $.

For any fill-empty profile $ \epsilon $, we have,
\begin{eqnarray}
n_{\epsilon, T}(T) 
& = & 
n_{\epsilon, T}(t_{\epsilon_0}) + r_f T_f(I_2; \epsilon) - r_e T_e(I_2; \epsilon) 
\nonumber \\
& = &
r_f T_f(I_1; \epsilon) - r_e T_e(I_1; \epsilon) 
\nonumber \\
&   &
+
r_f T_f(I_2; \epsilon) - r_e T_e(I_2; \epsilon)
\end{eqnarray}  
Since $ T_f(I_2; \epsilon) + T_e(I_2; \epsilon) \leq T - t_{\epsilon_0} $, we have,
\begin{eqnarray}
0
& \leq & 
n_{\epsilon, T}(T) 
\nonumber \\
& \leq & 
r_f (T_f(I_1; \epsilon_0) - T_{e, 1}(I_1; \epsilon)) - r_e T_e(I_1; \epsilon) 
\nonumber \\
&   &
+
r_f (T - t_{\epsilon_0} - T_e(I_2; \epsilon)) - r_e T_e(I_2; \epsilon)
\end{eqnarray}
which may be re-arranged to give,
\begin{eqnarray}
T_e(I_2; \epsilon) 
& \leq &
\frac{r_f T_f(I_1; \epsilon_0) - r_e T_e(I_1; \epsilon)}{r_e} 
\nonumber \\
&   &
- 
\frac{r_f}{r_f + r_e} (T_e(I_1; \epsilon_0) + T_{e, 1}(I_1; \epsilon) - T_e(I_1; \epsilon))
\nonumber \\
\end{eqnarray} 
where the derivation of this inequality makes use of the identity
$ r_f T_f(I_1; \epsilon_0) - r_e T_e(I_1; \epsilon_0) = r_e (T - t_{\epsilon_0}) $.

We then obtain that,
\begin{eqnarray}
n_{\epsilon, P}(T) 
& = &
r_e (T_e(I_1; \epsilon) + T_e(I_2; \epsilon)) 
\nonumber \\
& \leq &
r_f T_f(I_1; \epsilon_0) 
\nonumber \\
&   &
- \frac{r_f r_e}{r_f + r_e}
(T_e(I_1; \epsilon_0) + T_{e, 1}(I_1; \epsilon) 
\nonumber \\
&   &
- T_e(I_1; \epsilon))
\end{eqnarray}
Since $ n_{\epsilon_0, P}(T) = r_f T_f(I_1; \epsilon_0) $, then
if we can show that $ T_e(I_1; \epsilon_0) + T_{e, 1}(I_1; \epsilon) -
T_e(I_1; \epsilon) \geq 0 $, we will have established that
$ n_{\epsilon, P}(T) \leq n_{\epsilon_0, P}(T) $, thereby proving
the maximality of $ \epsilon_0 $. 

So, suppose $ T_e(I_1; \epsilon_0) + T_{e, 1}(I_1; \epsilon) -
T_e(I_1; \epsilon) < 0 $.  Then defining $ I(t) = [0, t] $,
we may note that the function $ \tau(t) \equiv T_e(I(t); \epsilon_0) 
+ T_{e, 1}(I(t); \epsilon) - T_e(I(t); \epsilon) $ is continuous,
and satisfies $ \tau(0) = 0 $, $ \tau(t_{\epsilon_0}) < 0 $.  Let
us then define $ t^* = \inf \{t \in [0, t_{\epsilon_0}]| \tau(t) < 0\} $.
By continuity of $ \tau $ and from the definition of $ \inf $, we have
that $ t^* < t_{\epsilon_0} $, $ \tau(t^*) = 0 $, and that for any $ t > t^* $ there exists 
a $ t' \in (t^*, t) $ such that $ \tau(t') < 0 $.

Now, by assumption $ \delta(t) $ is piecewise constant, hence
if $ \delta $ is discontinuous at $ t^* $, then there exists
an interval $ (t^*, t^* + h) \subset I_1 $ over which $ \delta $ is constant.
If $ \delta $ is continuous at $ t^* $, then there also exists
an interval $ (t^*, t^* + h) \subset I_1 $ over which $ \delta $ is
constant.  

Suppose $ \delta(t) = 1 $ on $ (t^*, t^* + h) $.  Then for any
$ h' < h $, we have that $ T_e(I(t^* + h'); \epsilon_0) =
T_e(I(t^*); \epsilon_0) $, since the prescription for 
$ \epsilon_0 $ is to fill when $ \delta(t) = 1 $ on $ I_1 $.
We also have that $ T_{e, 1}(I(t^* + h'); \epsilon) -
T_e(I(t^* + h'); \epsilon) = T_{e, 1}(I(t^*); \epsilon) -
T_e(I(t^*); \epsilon) + T_{e, 1}([t^*, t^* + h']; \epsilon) -
T_e([t^*, t^* + h']; \epsilon) $.  Since $ \delta(t) = 1 $ 
on $ (t^*, t^* + h') $, it follows that $ T_{e, 1}([t^*, t^* + h'];
\epsilon) = T_e([t^*, t^* + h']; \epsilon) $, and hence that
$ T_{e, 1}(I(t^* + h'); \epsilon) - T_e(I(t^* + h'); \epsilon) =
T_{e, 1}(I(t^*); \epsilon) - T_e(I(t^*); \epsilon) $.  Therefore,
$ \tau(t^* + h') = \tau(t^*) $, so that $ \tau(t) = 0 $ on
$ [t^*, t^* + h] $, contradicting the definition of $ t^* $.

So, suppose $ \delta(t) = 0 $ on $ (t^*, t^* + h) $.  Then
it should be clear that $ T_{e, 1}(I(t); \epsilon) $ is
constant over $ (t^*, t^* + h) $.  If $ n_{\epsilon_0, T}(t^*) > 0 $,
then according to our prescription there exists an $ h' \in (0, h) $
such that $ \epsilon_e(t) = 0 $ with $ n_{\epsilon_0, T}(t) > 0 $
over $ (t^*, t^* + h') $.  Therefore, given $ h'' \in (0, h') $, we have
$ T_e(I(t^* + h''); \epsilon_0) = T_e(I(t^*); \epsilon_0) + h'' $, while 
$ T_e(I(t^* + h''); \epsilon) = T_e(I(t^*); \epsilon) + T_e([t^*, t^* + h'']; \epsilon) $.
The result is that $ \tau(t^* + h'') = \tau(t^*) + h'' - T_e([t^*, t^* + h'']; \epsilon)
\geq \tau(t^*) = 0 $.  Therefore, $ \tau(t) \geq 0 $ on $ [t^*, t^* + h'] $,
again contradicting the definition of $ t^* $.

So, suppose that $ \delta(t) = 0 $ on $ (t^*, t^* + h) $ with
$ n_{\epsilon_0, T}(t^*) = 0 $.  Then,
\begin{eqnarray} 
n_{\epsilon, T}(t^*) 
& = &
r_f T_f(I(t^*); \epsilon) - r_e T_e(I(t^*); \epsilon) 
\nonumber \\
& \leq &
r_f (T_f(I(t^*); \epsilon_0) - T_{e, 1}(I(t^*); \epsilon)) 
\nonumber \\
&   &
-
r_e (T_e(I(t^*); \epsilon_0) + T_{e, 1}(I(t^*); \epsilon)) 
\nonumber \\
& = &
n_{\epsilon_0, T}(t^*) - (r_f + r_e) T_{e, 1}(I(t^*); \epsilon)
\nonumber \\
& = &
-(r_f + r_e) T_{e, 1}(I(t^*); \epsilon)
\end{eqnarray}
which is only possible if $ n_{\epsilon, T}(t^*) = 0 $ 
with $ T_{e, 1}(I(t^*); \epsilon) = 0 $.  But, since
$ n_{\epsilon, T}(t^*) = n_{\epsilon_0, T}(t^*) = 0 $,
then since $ \delta(t) = 0 $ on $ (t^*, t^* + h) $,
it follows that $ n_{\epsilon, T}(t) = n_{\epsilon_0, T}(t) = 0 $
on $ (t^*, t^* + h) $, and hence $ T_{e, 1}([t^*, t^* + h']; \epsilon) = 
T_e([t^*, t^* + h']; \epsilon) = T_e([t^*, t^* + h']; \epsilon_0) = 0 $
on for all $ h' \in [0, h] $, so that $ \tau(t) = \tau(t^*) = 0 $
on $ [t, t + h] $, which is again a contradiction.

Since we have exhausted all possibilites, we have established
that $ \tau(t_{\epsilon_0}) < 0 $ leads to a contradiction.
Therefore, $ \tau(t_{\epsilon_0}) \geq 0 $, and the proof
is complete.

We should note that, although the optimal fill-empty profile 
$ \epsilon_0 $ may not necessarily be unique, if
$ \epsilon $ denotes any other optimal fill-empty
profile, then we must have $ T_e([0, T]; \epsilon) =
T_e([0, T]; \epsilon_0) $.  It may be readily shown
that $ n_{\epsilon, T}(T) = n_{\epsilon_0, T}(T) = 0 $:

An $ \epsilon $ for which $ n_{\epsilon, T}(T) > 0 $ is not
optimal, for letting $ t_{\epsilon} $ denote when $ n_{\epsilon, T}(t) =
r_e (T - t) $ (by the Intermediate Value Theorem, such a $ t $ exists),  
we have $ n_{\epsilon, P}(T) = n_{\epsilon, P}(t_{\epsilon}) + 
r_e T_e([t_{\epsilon}, T]; \epsilon) \leq n_{\epsilon, P}(t_{\epsilon}) +
r_e (T - t_{\epsilon}) $, with equality only occurring when 
$ T_e([t_{\epsilon}, T]; \epsilon) = r_e (T - t_{\epsilon}) $.  
However, $ T_e([t_{\epsilon}, T]; \epsilon) = r_e (T - t_{\epsilon}) $
implies that $ n_{\epsilon, T}(T) = 0 $, and so our claim is proved.

But this implies that $ r_f T_f([0, T]; \epsilon) =
r_e T_e([0, T]; \epsilon) = r_e T_e([0, T]; \epsilon_0) =
r_f T_f([0, T]; \epsilon_0) $, so that $ T_f([0, T]; \epsilon)
= T_f([0, T]; \epsilon_0) $.  Finally, $ T_w([0, T]; \epsilon) =
T - T_f([0, T]; \epsilon) - T_e([0, T]; \epsilon) =
T - T_f([0, T]; \epsilon_0) - T_e([0, T]; \epsilon_0) =
T_w([0, T]; \epsilon_0) $.

Therefore, although the optimal fill-empty profile may not
be unique, the $ T_f $, $ T_e $, and $ T_w $ values are
uniquely specified.

\subsection{Examples of optimal fill-empty profiles}

For simplicity, we consider optimal fill-empty profiles generated
by a $ \delta(t) $ that is periodic over the time interval $ [0, T] $.
Specifically, we consider a basic profile, denoted $ \sigma_{(T_1, T_2)} $,
defined by,
\begin{eqnarray}
\sigma_{(T_1, T_2)}(t) =
\left\{ \begin{array}{cc}
  1 &
  \mbox{if $ t \in [0, T_1] $} \nonumber \\
  0 &
  \mbox{if $ t \in (T_1, T_1 + T_2] $}
  \end{array}
\right.
\end{eqnarray}
Over the time interval $ [0, T] $, we then define $ \delta(t) $ by
setting $ \delta(t) = \sigma_{(T_1, T_2)}(t) $ on $ [0, T_1 + T_2] $,
and then imposing the periodicity relation $ \delta(t) = \delta(t + T_1 + T_2) $.
We also assume that $ T = N (T_1 + T_2) $, for some positive integer $ N $.

Now, if $ r_e T_2 \geq r_f T_1 $, then it should be apparent that an optimal
fill-empty profile, defined by our $ \epsilon_0 $ prescription, is to fill
whenever $ \delta(t) = 1 $, and to empty whenever $ \delta(t) = 0 $ and
$ n_{\epsilon, T}(t) > 0 $, until time $ T $.  For this profile, then,
we have that the optimal values of $ T_f $, $ T_e $, and $ T_w $ are
given by,
\begin{eqnarray}
&   &
T_f([0, T]; \epsilon_0) = N T_1 \nonumber \\
&   &
T_e([0, T]; \epsilon_0) = \frac{r_f}{r_e} N T_1 \nonumber \\
&   &
T_w([0, T]; \epsilon_0) = N \frac{r_e T_2 - r_f T_1}{r_e}
\end{eqnarray}
and so any other optimal fill-empty profile must fill exactly
when $ \delta(t) = 1 $.  Note that $ T_w $ only occurs in intervals
where $ \delta(t) = 0 $ (since $ T_f([0, T]; \epsilon_0) = N T_1 $).
Therefore, during these periods, since there is a lack of available resource 
(analogous to periods of night when making the analogy to sleep), 
the tank is either filling when there is nothing to fill it with (analogous
to wakefulness periods during lack of available sensory information), or
is emptying when the tank is already empty (analogous to excess sleeping).
When $ \delta(t) = 0 $, any combination of filling, or emptying when $ n_T = 0 $, 
over these time intervals will not affect the final value of
$ n_P(T) $.  Thus, there is a considerable degree of freedom in choosing
an optimal fill-empty profile, which is consistent with the observed
breakdown in sleeping patterns in environments with limited exposure
to the sun.

Now consider the case when $ r_e T_2 < r_f T_1 $.  Then at the end of
every resource availability cycle $ [n (T_1 + T_2), (n + 1) (T_1 + T_2)] $,
we have that $ n_{\epsilon_0, T} $ increases by $ r_f T_1 - r_e T_2 $,
as long as we are before the final emptying phase.  So, if $ n $ is chosen
so that $ n (r_f T_1 - r_e T_2) \leq r_e (T - n (T_1 + T_2)) $ but
$ (n + 1) (r_f T_1 - r_e T_2) > r_e (T - (n + 1) (T_1 + T_2)) $, then we have
$ t_{\epsilon_0} \in [n (T_1 + T_2), (n + 1) (T_1 + T_2)] $, which
further implies that $ t_{\epsilon_0} \in [n (T_1 + T_2), n (T_1 + T_2) + T_1] $.
Therefore, $ n_{\epsilon, T}(t_{\epsilon_0}) = n (r_f T_1 - r_e T_2) +
r_f (t_{\epsilon_0} - n (T_1 + T_2)) = r_e (T - t_{\epsilon_0}) $.

Solving for $ t_{\epsilon_0} $, we obtain,
\begin{equation}
t_{\epsilon_0} = n T_2 + \frac{r_e}{r_e + r_f} T
\end{equation}
from which it follows that,
\begin{eqnarray}
&   &
T_f([0, T]; \epsilon_0) = \frac{r_e}{r_e + r_f} T
\nonumber \\
&   &
T_e([0, T]; \epsilon_0) = \frac{r_f}{r_e + r_f} T
\nonumber \\
&   &
T_w([0, T]; \epsilon_0) = 0
\end{eqnarray}

An interesting fill-empty profile that arises from this condition is defined
as follows:  Over the interval $ [0, T] $, $ \epsilon_e(t) = 1 $ whenever
$ \delta(t) = 0 $.  Whenever $ \delta(t) = 1 $, $ \epsilon_f(t) $ and
$ \epsilon_e(t) $ alternate in being $ 1 $, with corresponding time
lengths $ t_f $, $ t_e $.  To determine these time lengths, we first
assume that there are $ M $ such cycles over each $ \delta(t) = 1 $ interval of length
$ T_1 $, so that $ t_f + t_e = T_1/M $.  The net accumulation in the tank
over each such interval should be $ r_e T_2 = M (r_f t_f - r_e t_e) $,
which may be solved to give,
\begin{eqnarray}
&   &
t_f = \frac{1}{M} \frac{r_e}{r_f + r_e} (T_1 + T_2)
\nonumber \\
&   &
t_e = \frac{1}{M} \frac{r_f T_1 - r_e T_2}{r_f + r_e}
\end{eqnarray}
Note that $ r_f t_f > r_e t_e $, and that $ T_e([0, T]; \epsilon) =
r_f/(r_e + r_f) T $, so that this profile is indeed an optimal
one.

Essentially, when
the amount of time during which $ \delta(t) = 0 $ is not sufficiently
long to process all of the resource that can fill the tank, then
one optimal solution profile is, during periods when $ \delta(t) = 1 $,
to fill and empty the tank in alternating time intervals of length
$ t_f $, $ t_e $, and then to empty the tank whenever $ \delta(t) = 0 $.
We argue that the $ t_e $ empty periods are analogous to the phenomenon
of ``microsleep'' that occurs during sleep deprivation.

In general, we claim that any optimal solution profile for
this form of $ \delta(t) $ will have $ \epsilon_e(t) = 1 $ whenever 
$ \delta(t) = 0 $.  Otherwise, we obtain,
$ r_f/(r_f + r_e) T = T_e([0, T]; \epsilon) = 
T_{e, 0}([0, T]; \epsilon) + T_{e, 1}([0, T]; \epsilon) 
< N T_2 + T_{e, 1}([0, T]; \epsilon) \Rightarrow
T_{e, 1}([0, T]; \epsilon) > N (r_f T_1 - r_e T_2)/(r_f + r_e) $.

But this implies that $ T_f([0, T]; \epsilon) \leq N T_1 -
T_{e, 1}([0, T]; \epsilon) < N T_1 - N (r_f T_1 - r_e T_2)/(r_f + r_e)
= r_e/(r_e + r_f) T \Rightarrow\Leftarrow $.  With this
contradiction, our claim is proved. 

As a final example for this subsection, we consider a resource
availability profile given by,
\begin{widetext}
\begin{eqnarray}
\delta(t) = 
\left\{ \begin{array}{cc}
  \sigma_{(T_1, T_2)}(t) &
  \mbox{if $ t \in [0, T_1 + T_2] $} \nonumber \\
  \sigma_{(T_1, T_2')}(t + T_1 + T_2) &
  \mbox{if $ t \in [T_1 + T_2, T = 2 T_1 + T_2 + T_2'] $}
  \end{array}
\right.
\end{eqnarray}
\end{widetext}
where $ r_e T_2' < r_f T_1 < r_e T_2 $.  Then it is possible to show that,
\begin{eqnarray}
&   &
T_f([0, T]; \epsilon_0) = T_1 + \frac{r_e}{r_e + r_f} (T_1 + T_2')
\nonumber \\
&   &
T_e([0, T]; \epsilon_0) = \frac{r_f}{r_e} T_1 + \frac{r_f}{r_e + r_f} (T_1 + T_2')
\nonumber \\
&   &
T_w([0, T]; \epsilon_0) = \frac{r_e T_2 - r_f T_1}{r_e}
\end{eqnarray}

We now show that this uniquely determines $ T_f([0, T_1 + T_2]; \epsilon_0) $,
$ T_f([T_1 + T_2, T]; \epsilon_0) $, $ T_e([0, T_1 + T_2]; \epsilon_0) $,
$ T_e([T_1 + T_2, T]; \epsilon_0) $, $ T_w([0, T_1 + T_2]; \epsilon_0) $,
$ T_w([T_1 + T_2, T]; \epsilon_0) $.
 
Note that $ T_f([0, T]; \epsilon_0) \geq T_f([0, T_1 + T_2]; \epsilon_0) +
\frac{r_e}{r_e + r_f} (T_1 + T_2') \Rightarrow T_f([T_1 + T_2, T]; \epsilon_0) \geq
\frac{r_e}{r_e + r_f} (T_1 + T_2') $.  Therefore, $ T_e([T_1 + T_2, T]; \epsilon_0)
\leq \frac{r_f}{r_e + r_f} (T_1 + T_2') $.  However, $ n_{\epsilon_0, T}(T) =
0 \geq r_f T_f([T_1 + T_2, T]; \epsilon_0) - r_e T_e([T_1 + T_2, T]; \epsilon_0)
\geq \frac{r_f r_e}{r_e + r_f} (T_1 + T_2') - r_e T_e([T_1 + T_2, T]; \epsilon_0)
\Rightarrow T_e([T_1 + T_2, T]; \epsilon_0) \geq \frac{r_f}{r_e + r_f} (T_1 + T_2') $,
and so $ T_e([T_1 + T_2, T]; \epsilon_0) = \frac{r_f}{r_e + r_f} (T_1 + T_2') $.
But this implies that $ T_f([T_1 + T_2, T]; \epsilon_0) \leq \frac{r_e}{r_e + r_f} (T_1 + T_2') $,
so $ T_f([T_1 + T_2, T]; \epsilon_0) = \frac{r_e}{r_e + r_f}(T_1 + T_2') $.  This
further implies that $ T_w([T_1 + T_2, T]; \epsilon_0) = 0 $, and so
that $ T_w([0, T_1 + T_2]; \epsilon_0) = \frac{r_e T_2 - r_f T_1}{r_e} $.
Finally, we obtain $ T_f([0, T_1 + T_2]; \epsilon_0) = T_1 $, and
$ T_e([0, T_1 + T_2]; \epsilon_0) = \frac{r_f}{r_e} T_1 $.

Summarizing the results, we have,
\begin{eqnarray}
&   &
T_f([0, T_1 + T_2]; \epsilon_0) = T_1
\nonumber \\
&   &
T_f([T_1 + T_2, T]; \epsilon_0) = \frac{r_e}{r_e + r_f} (T_1 + T_2') < T_1
\nonumber \\
&   &
T_e([0, T_1 + T_2]; \epsilon_0) = \frac{r_f}{r_e} T_1
\nonumber \\
&   &
T_e([T_1 + T_2, T]; \epsilon_0) = \frac{r_f}{r_f + r_e}(T_1 + T_2') > T_2'
\nonumber \\
&   &
T_w([0, T_1 + T_2]; \epsilon_0) = \frac{r_e T_2 - r_f T_1}{r_e}
\nonumber \\
&   &
T_w([T_1 + T_2, T]; \epsilon_0) = 0
\end{eqnarray}
Therefore, maximal processing of external resource still requires emptying the
tank when $ \delta(t) = 1 $ on the interval $ [T_1 + T_2, T] $.  In making
the analogy with sleep, this implies that excess sleep in a certain time interval
will not prevent sleep deprivation in a later time interval when the period of
resource availability exceeds the period when the resource may be processed.
Intuitively, this makes sense, since, once the brain has ``taken out the trash,''
further sleep will not prevent the accumulation of ``trash'' during a later cycle.

\section{Conclusions and Future Research}

This paper presented a simplified model of the filling and emptying
of a tank as an analogy with which to analyze the phenomenon of sleep.
Our model is motivated by a hypothesis that sleep is the brain's way
of processing material accumulated during periods of sensory input.
By analogy with a corporation, which may need to periodically update 
inventories and finances for proper functioning, so too the brain may
need to periodically cut-off external inputs and process accumulated
material, in order to maintain updated information with which to
properly analyze the environment.

Based on our tank filling model, we were able to show that 
when there are alternating periods of availability of
external inputs, the optimal amount of information can be
processed if information is collected when it is available 
(during the day for the vast majority of organisms),
and processed when it is not, or considerably less, available
(at night, again, for the vast majority of organisms).
We showed that when there are excessive periods where
resource is unavailable, then there is a continuum of optimal 
fill-empty profiles, involving either filling when no resource is 
present, or emptying when the tank is empty.  Thus, our model is consistent with the
observed breakdown in sleeping patterns in environments with
minimal exposure to the sun.

Furthermore, we showed that when the periods of information deficiency are
not sufficiently long to process all of the collected information,
an optimal information collection strategy can involve periodic 
tank emptying intervals during periods of information availability.
We argued that this behavior is analogous to the observed
phenomenon of ``microsleeps.''

Our model also suggests that it is not possible
to store sleep time beyond a certain amount, so that periods
of extended sleep will not prevent sleep deprivation at a later
time.  Our model also captures the breakdown in sleeping patterns

Although our model is consistent with a number of phenomena
associated with sleep, it is nevertheless highly simplistic:
Our model assumes that the tank can be filled
without limit, corresponding to an organismal ability to
maintain unlimited periods of wakefulness.  This of course
is incorrect, because an over-accumulation of unprocessed
information will eventually lead to improper brain function.  
Therefore, one important modification that
must be made to our model is to place an upper bound
on $ n_T $, denoted $ n_{T, max} $, so that the fill-empty profile must be
such that $ n_{\epsilon, T}(t) \leq n_{T, max} $ at all times.

Furthermore, our model also assumes that there is no time
delay associated with switching tasks (from fill to empty
and vice versa).  In reality, when task switching occurs,
there are generally transients associated with the time 
it takes for the components for performing one task to shut-down,
and for another set of components for performing the other task
to start up.  These transients lead to a time cost $ \tau_{switch} $
during which neither system is active, and so future models
will need to consider positive values of $ \tau_{switch} $
when determining optimal fill-empty profiles.

In addition, pursuing the analogy with a corporation, we may
note that although inventories and finances may be updated at
regular intervals, it may nevertheless be an optimal strategy
to also perform some level of upkeep during periods of external
input.  A lower level of upkeep could have shorter associated
transients, and by lessening the rate of accumulation of
material, will allow for a greater overall input of external
resource before system-wide upkeep becomes necessary.

Our model considered the filling and emptying of
a single tank.  In reality, an organ such as the brain consists
of billions of interacting neurons, and thus our model should
be extended to include interactions amongst numerous tanks.
An important issue to consider is neuronal synchronization,
whereby the neurons oscillate between their waking and
sleep modes together.  Presumably, since it is advantageous
for each component of a system to switch between input and processing
tasks in accordance with external resource availability,
and since system performance is optimized when all components
coordinate their actions, the strategy whereby all system
components collectively switch between input and processing tasks in
accordance with external resource availability is a self-reinforcing
one.

As a final remark, because sleep arises as the need for the 
brain to perform upkeep tasks that accumulate over a period
of sensory input, mathematical models related to sleep could
also be useful for analyzing other interconnected structures
where a proper ordering of tasks could lead to optimal
system performance.  In particular, models of sleep could
be useful in understanding systems that exhibit oscillatory
periods of relatively intense apparent activity (as exhibited
by absorption of information or material into the system), followed
by periods of apparent stasis.

\begin{acknowledgments}

This research was supported by the Israel Science Foundation.

\end{acknowledgments}

\end{document}